# *SbD*-Based Synthesis of Low-Profile *WAIM* Superstrates for Printed Patch Arrays

G. Oliveri, *Senior Member, IEEE*, A. Polo, *Member, IEEE*, M. Salucci, *Member, IEEE*, G. Gottardi, *Member, IEEE*, and A. Massa, *Fellow, IEEE*

*Abstract*—An innovative strategy for the design of low-profile wide-angle impedance matching (*WAIM*) superstrates to active return loss (*ARL*) stability of scanned microstrip-printed periodic phased arrays (*PPAs*) is presented. Towards this end, an instance of the *System-by-Design* (*SbD*) paradigm is exploited and a task-oriented formulation of the *WAIM* synthesis problem is proposed. More specifically, a customized global optimization strategy is combined with an accurate and fast modeling of the array *ARL*, which is based on a semi-analytical approach, to enable an effective design of multi-layer structures with both isotropic and anisotropic materials. A set of representative numerical examples, concerned with different microstrip patch substrates, lattices, and design constraints, is presented to give the interested readers some insights on the features and the potentialities of the proposed implementation.

## I. INTRODUCTION AND RATIONALE

**W**IDE-ANGLE electronic scanning is a feature required to planar phased arrays in an increasing number of applicative scenarios including terrestrial/satellite communications, automotive radars, and remote sensing systems [1]-[8]. The main motivation is that covering a wide field-of-view (*FoV*) with a single flat antenna system enables considerable architectural simplifications and non-negligible cost savings with respect to alternative solutions based, for instance, on mechanically scanned and/or faceted/conformal arrays [1][4][5]. However, achieving acceptable performance in a wide *FoV* is often a challenging task, especially in large arrays, because of the variations of both the array realized gain and the active return loss (*ARL*) that may cause severe scan losses as well as

Manuscript received September 0, 2019

This work has been partially supported by the Italian Ministry of Education, University, and Research within the Program "Smart cities and communities and Social Innovation" (CUP: E44G14000060008) for the Project "WATERTECH - Smart Community per lo Sviluppo e l'Applicazione di Tecnologie di Monitoraggio Innovative per le Reti di Distribuzione Idrica negli usi idropotabili ed agricoli" (Grant no. SCN_00489) and within the Program PRIN 2017 for the Project "CLOAKING METASURFACES FOR A NEW GENERATION OF INTELLIGENT ANTENNA SYSTEMS (MANTLES)".

G. Oliveri, A. Polo, M. Salucci, G. Gottardi, and A. Massa are with the ELEDIA Research Center (ELEDIA@UniTN - University of Trento), Via Sommarive 9, 38123 Trento - Italy (e-mail: {giacomo.oliveri, alessandro.polo.1, marco.salucci, giorgio.gottardi, andrea.massa}@unitn.it)

A. Massa is also with the ELEDIA Research Center (ELEDIA@UESTC - UESTC), School of Electronic Engineering, Chengdu 611731 - China (e-mail: andrea.massa@uestc.edu.cn)

A. Massa is also with the ELEDIA Research Center (ELEDIA@TSINGHUA - Tsinghua University), 30 Shuangqing Rd, 100084 Haidian, Beijing - China (e-mail: andrea.massa@tsinghua.edu.cn)

blind spot effects [9]-[14]. As for this latter, different concepts have been proposed to mitigate/avoid *ARL* stability issues in the state-of-the-art literature [9][12]-[15]. For instance, techniques based on the *mutual coupling cancellation* have been introduced to yield an indirect minimization of the *ARL* variations [9][15]. By combining the antenna array with a *decoupling* feeding network, designed to counterbalance/cancel the coupling between any pair of adjacent antennas, the active element impedance is kept as close as possible to the passive one regardless of the scan angle [9][15]. Notwithstanding, such a solution is more suitable for narrowband applications [15] because of the design/fabrication difficulties, the power consumption, and the costs of a wideband multi-element decoupling network.

Alternatively, the exploitation of wide-angle impedance matching (*WAIM*) superstrates, based on layered dielectric/metallic materials, has been presented in [10][12]-[14] for *ARL* stabilization. The success of such an approach, proved in different applicative scenarios, is mainly due to the possibility of obtaining wide *FoV*s and adequate bandwidths with low-profile structures and low-cost implementations [10][13] also in printed technology [12][13] by adopting rules and methods drawn from the metamaterial engineering [16][17]. On the other hand, the design of *WAIM*s with standard iterative methodologies needs the full-wave modeling of the ensemble of the array element and the superstrate arrangement [18], which often implies numerically cumbersome procedures [12][13][18]. As a consequence, state-of-the-art *WAIM* design approaches exploiting commercial full-wave electromagnetic modeling softwares rely on gradient-based local search techniques [18]. Otherwise and only for canonical truncated waveguide antenna arrays [13][14], computationally-feasible and effective *WAIM* layouts have been designed which considerably enhance *ARL* stability thanks to the exploitation of hybrid analytical models within the System-by-Design (*SbD*) framework [19][20]. However, no solution has been proposed so far to generalize such an approach to other arrays such as the printed ones despite (*i*) their widespread application and relevance in next-generation communications and radar antennas, (*ii*) their low fabrication costs and low profile, and (*iii*) the inherent *ARL* instability issues of such class of radiators caused by the unavoidable antenna coupling and related to the surface waves propagating within the substrate. In this paper, the technique introduced in [13] is extended and customized to the design of low-profile *WAIM* multi-layer dielectric superstrates for enhancing the scan range of microstrip-printed periodic phased arrays (*PPAs*). More specifically, the stabilization of the *ARL* is addressed, likewise [12][13][14], owing to its fundamental importance to avoid performance deterioration and, indirectly, realized gain reductions in phased array feed structures during scanning. To achieve this goal, a semi-analytical description of the *ARL* of *PPA*s, when coated with multi-layer *WAIM* superstrates, is derived and combined with an efficient optimization strategy to define, on the one hand, (*i*) an accurate, reliable, and fast physical model of the whole device (i.e., *PPA* and *WAIM* layer [*PPAW*]) and, on the other hand, (*ii*) a synthesis method based on the effective exploration of the space of physically-admissible solutions. The infinite array approximation [1] is employed for this purpose, likewise in [11][12][13][14][18], owing to its well-known efficiency and accuracy when dealing with large-array scenarios [1].

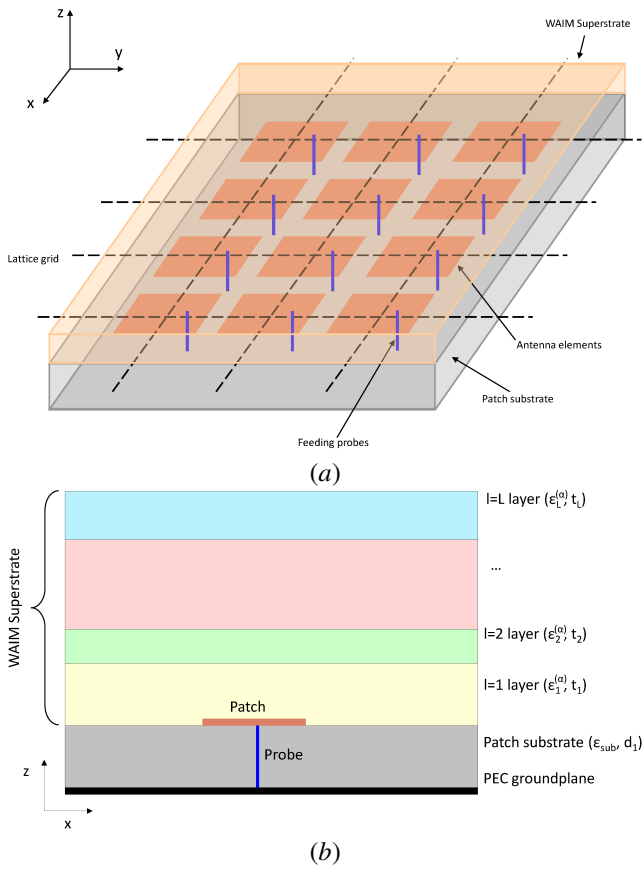

Figure 1.   *Problem Geometry*. Sketch of (*a*) the *WAIM*-coated phased array and (*b*) the side-view of the elementary radiator.

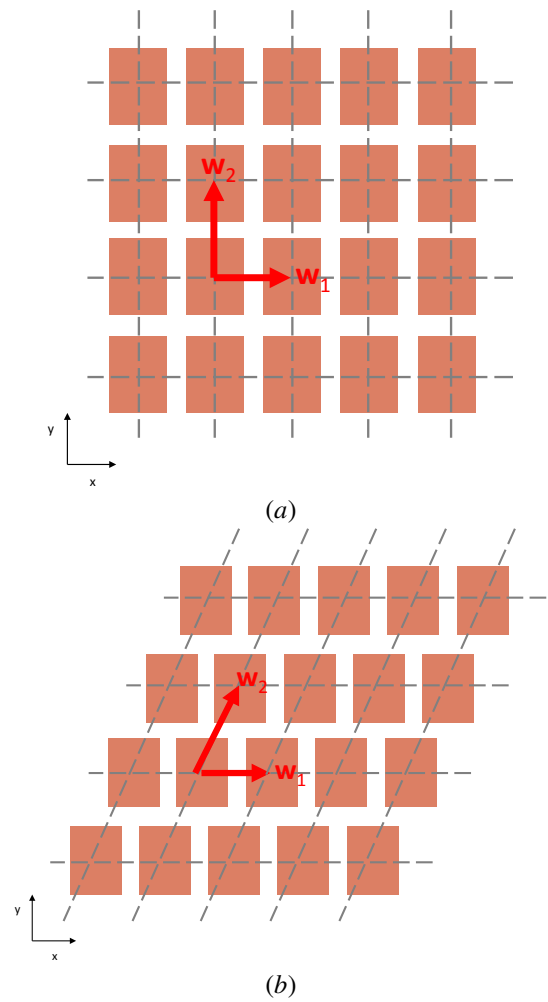

Figure 2.   *Problem Geometry*. Top-view of the *WAIM*-coated layout when arranged according to (*a*) a square or (*b*) a triangular lattice.

The methodological foundations of such an approach are the following ones: (*i*) *SbD*-based synthesis strategies have already demonstrated their effectiveness in solving *WAIM* design problems formulated in a task-oriented fashion [13][14], (*ii*) modeling the array *ARL* in a semi-analytical fashion is expected to guarantee very efficient and accurate computations [11] without cumbersome full-wave simulations or training of surrogate approximate models, (*iii*) global optimization techniques are suitable countermeasures when dealing with multi-minima and multi-scale synthesis problems.

From a methodological perspective, the main innovative contributions of this work lie in (*i*) the reformulation of the method in [13] to the design of *WAIM* low-profile radomes to be combined with *PPA*s, (*ii*) the derivation of a semi-analytical model for the computation of the *ARL* for a generic array lattice and multi-layer superstrate geometries, patch dimensions, and feed probe positions, (*iii*) the extension of the approach in [11] from isotropic to uniaxial dielectric layers, and (*iv*) the derivation of suitable design guidelines for the *SbD*-based synthesis strategy of *PPAW*s (i.e., the definition, the implementation, and the demonstration of the suitable tools and methodological approaches to be integrated in the *SbD* paradigm for the design of effective *WAIM*s for *PPA*s).

The outline of the paper is as follows. First, the task-oriented synthesis problem is formulated (Sect. II), then the solution strategy, featuring an instance of the *SbD* paradigm, is illustrated in Sect. III. A set of representative examples, drawn from an extensive numerical validation, is discussed (Sect. IV) to assess the effectiveness and the efficiency of the proposed design method when applied to arrays with different lattices, substrates, and antenna sizes. Finally, some conclusions follow (Sect. V).

## II. Task-Oriented Synthesis Problem Statement

The synthesis of a low-profile multi-layer *WAIM* superstrate [Fig. 1(*a*)] to enhance the performance stability of a *PPA* when scanning requires the maximization of the integral of the *ARL* across all steering angles and operative frequencies in order to avoid/mitigate the potential deterioration to the feed structures and, indirectly, realized gain reductions [13]. With reference to the geometry in Fig. 1(*a*), this problem can be mathematically formulated within the *SbD* framework as the following task-oriented one

> *Task-oriented SbD WAIM Synthesis Problem* - Find the optimal values of the vectors of the *WAIM* descriptors, $\mathbf{t}^{opt}$ and $\boldsymbol{\varepsilon}^{opt}$, such that

$$\left\{ \mathbf{t}^{opt}, \boldsymbol{\varepsilon}^{opt} \right\} = arg \left\{ \min_{\mathbf{t} \in \mathcal{T}, \boldsymbol{\varepsilon} \in \mathcal{E}} \left[ \Psi \left( \mathbf{t}, \boldsymbol{\varepsilon}, \mathbf{d} \right) \right] \right\} \quad (1)$$

where

$$\Psi \left( \mathbf{t}, \boldsymbol{\varepsilon}, \mathbf{d} \right) \triangleq \frac{1}{\int_{f_{\min}}^{f_{\max}} \psi \left( f; \mathbf{t}, \boldsymbol{\varepsilon}, \mathbf{d} \right) df} \quad (2)$$

is the task-oriented cost function, while the entries of $\mathbf{t}$ ($\mathbf{t} \triangleq \{t_l, l = 1, .., L\}$) and $\boldsymbol{\varepsilon}$ ($\boldsymbol{\varepsilon} \triangleq \left\{ \varepsilon_l^{(\alpha)}, l = 1, .., L, \alpha \in \{xx, yy, zz\} \right\}$) are the thickness of the *l*-th ($l = 1, ..., L$; $L$ being the number of layers of the *WAIM*) layer, $t_l$, and the components of the relative permittivity tensor $\overline{\overline{\varepsilon}}_l \triangleq \varepsilon_l^{(xx)} \widehat{\mathbf{x}} \widehat{\mathbf{x}} + \varepsilon_l^{(yy)} \widehat{\mathbf{y}} \widehat{\mathbf{y}} + \varepsilon_l^{(zz)} \widehat{\mathbf{z}} \widehat{\mathbf{z}}$ of the corresponding uniaxially anisotropic and non-magnetic material, $\left\{ \varepsilon_l^{(\alpha)}; \alpha \in \{xx, yy, zz\} \right\}$, [Fig. 1(*b*)], $\mathcal{T}$ and $\mathcal{E}$ being the associated feasibility sets that encode the fabrication/technological constraints. Moreover, the term

$$\psi \left( f; \mathbf{t}, \boldsymbol{\varepsilon}, \mathbf{d} \right) = \int_{\theta_{\min}}^{\theta_{\max}} \int_{\varphi_{\min}}^{\varphi_{\max}} ARL \left( \theta, \varphi, f; \mathbf{t}, \boldsymbol{\varepsilon}, \mathbf{d} \right) d\varphi d\theta \quad (3)$$

is the *integral active return loss* at the frequency $f$ and $ARL \left( \theta, \varphi, f; \mathbf{t}, \boldsymbol{\varepsilon}, \mathbf{d} \right)$ is the *ARL* seen at the antenna input port when the infinite array of identical rectangular patches with descriptors $\mathbf{d}$ is steered along the angular direction $(\theta, \varphi)$, $[f_{\min}, f_{\max}]$, $[\theta_{\min}, \theta_{\max}]$ and $[\varphi_{\min}, \varphi_{\max}]$ being the *WAIM* target bandwidth (to be specified within the antenna resonance band) and scan ranges in elevation and azimuth, respectively. The entries of the vector $\mathbf{d}$ ($\mathbf{d} \triangleq \{d_i, i = 1, .., I\}$) are the patch substrate thickness [$d_1$ - Fig. 1(*b*)] and its relative permittivity $\varepsilon_{sub}$ [$d_2 = \varepsilon_{sub}$ - Fig. 1(*b*)], the patch width, $d_3$, and its length, $d_4$, the probe feed offset along $x$, $d_5$, and $y$, $d_6$, and the array lattice vectors in the $x - y$ plane (i.e., $\mathbf{w}_1 \triangleq d_7 \widehat{\mathbf{x}} + d_8 \widehat{\mathbf{y}}$, $\mathbf{w}_2 \triangleq d_9 \widehat{\mathbf{x}} + d_{10} \widehat{\mathbf{y}}$ - Fig. 2). Accordingly, (3) accounts for the scan loss and gain variation effects caused by the varying impedance mismatch at the array element/feed network interface and the consequent radiated power losses when the antenna is steered [12], which is due to the interactions among the active elementary radiators [12]. Additional terms taking into account gain fluctuations and efficiency may be also seamlessly handled in the same framework by suitably extending (2). For instance, the maximization of the realized gain $RG \left( \theta, \varphi, f; \mathbf{t}, \boldsymbol{\varepsilon}, \mathbf{d} \right)$ may be accounted for by defining a new cost function as $\Psi' \left( \mathbf{t}, \boldsymbol{\varepsilon}, \mathbf{d} \right) \triangleq \left[ \int_{f_{\min}}^{f_{\max}} \int_{\theta_{\min}}^{\theta_{\max}} \int_{\varphi_{\min}}^{\varphi_{\max}} RG \left( \theta, \varphi, f; \mathbf{t}, \boldsymbol{\varepsilon}, \mathbf{d} \right) d\varphi d\theta \right]^{-1}$. However, such a formulation would require the derivation of a suitable semi-analytical model for the computation of $RG \left( \cdot \right)$, which is beyond the scope of the current paper (see Sect. III).

It is here worthwhile noticing that (2) may be optimized also with respect to the entries of $\mathbf{d}$ (i.e., the lattice, the substrate, the patch geometry, and the feeding probe position) besides $\mathbf{t}$ and $\boldsymbol{\varepsilon}$, but only the *WAIM* descriptors will be considered as degrees-of-freedom (*DoFs*) of the problem at hand to give some precise insights on the potentialities of multi-layer superstrates for *ARL* improvement when combined with *PPA*s without recurring to complex radiators, ad-hoc substrates, or non-canonical lattices. As regards the practicality of the resulting approach in terms of the possibility to synthesize the desired permittivity values, it is worth remarking that (*i*) several state-of-the art synthesis methods have been proposed to achieve such a goal [12][16][21][22][23][24][25], and that (*ii*) the design process can be easily constrained to exploit only commercially available laminates by suitably defining the feasibility sets $\mathcal{T}$ and $\mathcal{E}$. Moreover, the deduced results are also useful as a guideline regarding the best achievable performance for the envisaged structures, likewise [13] for waveguide arrays.

Finally, let us point out that the synthesis problem at hand is distinct from the theoretical and practical viewpoint from those in [13][14] since (*i*) there the radiating elements were filled truncated waveguides with coupling effects considerably different from those of printed radiators [1][11], (*ii*) consequently, the exploitation of a Galerkin method in the spectral domain is required here to perform the modeling the array *ARL* in a semi-analytical fashion, while a mode-matching strategy was used in [13].

## III. *SbD*-Based Design Technique featuring Semi-Analytical Modeling Block

With reference to (1), it is immediate to notice

- the highly non-linear nature of (2) that can induce severe issues in the search procedure owing to the (possible) occurrence of local minima in the functional landscape [26][27];
- the multi-scale nature of the problem at hand because of the presence of the micro-scale layer descriptors (i.e., the vectors $\mathbf{t}$ and $\boldsymbol{\varepsilon}$) along with the macro-scale dimension of the array performance index, $ARL \left( \theta, \varphi, f; \mathbf{t}, \boldsymbol{\varepsilon}, \mathbf{d} \right)$.

Those features represent the fundamental challenges to be addressed in synthesizing *PPAW*s. Towards this end, an extension and generalization of the strategy introduced in [13] is taken into account by reformulating the solution process in an iterative *SbD*-inspired one featuring the modular blocks in Fig. 3: (*i*) a *solution space exploration* block, which is responsible for the iterative generation of the trial *DoF* configurations to yield the optimal setup of the *PPAW*s descriptors, $\{\mathbf{t}^{opt}, \boldsymbol{\varepsilon}^{opt}\}$, through suitable global search techniques; (*ii*) an *electromagnetic modeling* block, which is in charge of the evaluation of the phased array physical response [i.e., $ARL \left( \theta, \varphi, f; \mathbf{t}, \boldsymbol{\varepsilon}, \mathbf{d} \right)$] corresponding to each guess *WAIM* configuration generated during the search process; (*iii*) a *physical linkage* block, which is devoted to the assessment of the *quality* of each design with respect to the performance targets and synthesis constraints. It is well-known that the key advantage of such a *SbD* formulation lies in its modularity [19] that enables the adoption/customization of the most effective tool for each functional block instead of using a standard "brute force" approach to the whole complex problem.

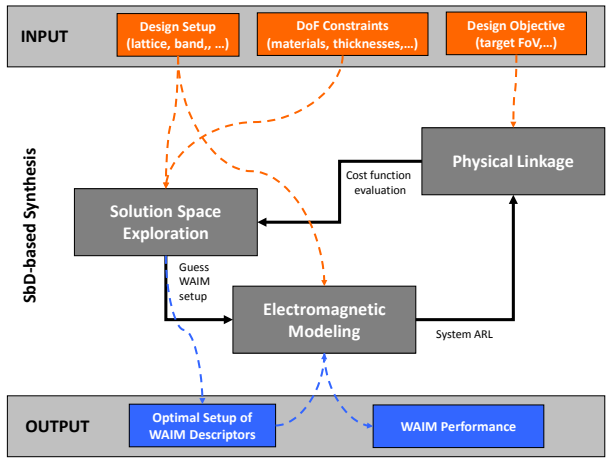

Figure 3. *SbD*-based Design Flowchart.

But what about the implementation of each sub-task? With reference to the *physical linkage* block and according to (1), it solely consists of the evaluation of the cost function (2) and the test of the compliance of the guess multi-layer superstrate configuration with the feasibility conditions encoded in $\mathcal{T}$ and $\mathcal{E}$.

According to the "No-free-lunch" theorems for optimization [26][27][28], the *solution space exploration* strategy is selected by taking into account both the properties of the cost function to be handled (2) and the type of unknowns and associated constraints within the search procedure. Due to the non-linear and multi-minima nature of $\Psi(\mathbf{t}, \boldsymbol{\varepsilon}, \mathbf{d})$, the (potential) non-convexity of the admissibility sets $\mathcal{T}$ and $\mathcal{E}$, and the continuous *DoF*s, an iterative approach based on the inertial-weight multi-agent particle swarm search procedure is employed [26][27][29][30]. More specifically, the global search process is performed by generating, at each $k$-th ($k = 1, ..., K$) iteration, a set of $R$ trial descriptors, $\mathbf{t}_r^{(k)}$ and $\boldsymbol{\varepsilon}_r^{(k)}$, $r = 1, ..., R$, starting from those at the $(k-1)$-th step that undergo the particle swarm *position update* and the *velocity update* operators ($\zeta_1$, $\zeta_2$, and $\zeta_3$ being the inertial weight, the cognitive acceleration, and the social acceleration coefficients [26]) subject to suitable boundary conditions [29][30]. The procedure is iterated until either $k = K$ ($K$ being the maximum number of iteration) or the stagnation condition

$$\left| \overline{\Psi}^{(\widetilde{K})} - \sum_{j=1}^{\omega} \frac{\overline{\Psi}^{(\widetilde{K}-j)}}{\mathcal{K}} \right| \leq \mathcal{C} \quad (4)$$

holds true at a given iteration $\widetilde{K}$ ($\widetilde{K} \in [1, K]$), $\omega$ and $\mathcal{C}$ being the window and threshold of convergence, respectively, while $\overline{\Psi}^{(\widetilde{K})}$ [$\overline{\Psi}^{(\widetilde{K})} \triangleq \min_{r=1,...,R;k=1,...,\widetilde{K}} \Psi(\mathbf{t}_r^{(k)}, \boldsymbol{\varepsilon}_r^{(k)}, \mathbf{d})$] is the value of the cost function (2) of the best solution found until the iteration $\widetilde{K}$. The final *WAIM* layout is then retrieved once the exploration of the functional landscape has been completed and it turns out to be

$$\{\mathbf{t}^{opt}, \boldsymbol{\varepsilon}^{opt}\} = arg \left\{ \min_{r=1,...,R;k=1,...,\min\{K,\widetilde{K}\}} \left[ \Psi\left(\mathbf{t}_r^{(k)}, \boldsymbol{\varepsilon}_r^{(k)}, \mathbf{d}\right) \right] \right\}, \quad (5)$$

that is the solution of (1). It is worth remarking that, unlike previous approaches exploiting gradient-based optimization strategies [18], a global multi-agent evolutionary approach is therefore considered in the *solution space exploration* phase. As for the last functional block (Fig. 3), the choice/customization of the most proper method to implement the *electromagnetic modeling* is a considerably more challenging task, from both the theoretical and the practical viewpoint, since a completely different formulation from that in [13] (which was based on a mode-matching strategy) is needed. As a matter of fact, the computation of the *ARL* for *PPA*s cannot be carried out according to the guidelines in [12][13] for open-ended waveguide arrangements because of the different mutual coupling phenomena [11][31]. In principle, the *PPAW*s can be modeled throughout full-wave electromagnetic solvers (as carried out in [18]), but this would result in a very computationally-expensive (sometimes almost or fully unpractical or unfeasible) process if a multi-agent global search technique is employed since a simulation should run for each guess solution generated within the *SbD* loop. Otherwise, an *ad-hoc* semi-analytical technique, inspired by [11] and extended to uniaxial dielectric layers, is here proposed. It exploits the Galerkin method in the spectral domain to compute the multi-layer Green's functions as detailed in the following.

Let us refer to the *ARL* of the *WAIM*-coated infinite array of identical probe-fed rectangular patches[1], which is defined according to [32] as

$$ARL(\theta, \varphi, f; \mathbf{t}, \boldsymbol{\varepsilon}, \mathbf{d}) \triangleq$$
$$\left| \frac{Z(0,0,f; \mathbf{t},\boldsymbol{\varepsilon},\mathbf{d}) + Z(\theta,\varphi,f; \mathbf{t},\boldsymbol{\varepsilon},\mathbf{d})}{Z(0,0,f; \mathbf{t},\boldsymbol{\varepsilon},\mathbf{d}) - Z(\theta,\varphi,f; \mathbf{t},\boldsymbol{\varepsilon},\mathbf{d})} \right|^2 \quad (6)$$

where $Z(\theta, \varphi, f; \mathbf{t}, \boldsymbol{\varepsilon}, \mathbf{d})$ is the array active input impedance given by [11]

$$Z(\theta, \varphi, f; \mathbf{t}, \boldsymbol{\varepsilon}, \mathbf{d}) \triangleq$$
$$-\sum_{m=1}^{M} C_m(\theta, \varphi, f; \mathbf{t}, \boldsymbol{\varepsilon}, \mathbf{d}) V_m(\theta, \varphi, f; \mathbf{t}, \boldsymbol{\varepsilon}, \mathbf{d}), \quad (7)$$

the terms $V_m(\theta, \varphi, f; \mathbf{t}, \boldsymbol{\varepsilon}, \mathbf{d})$ and $C_m(\theta, \varphi, f; \mathbf{t}, \boldsymbol{\varepsilon}, \mathbf{d})$ being the $m$-th mode voltage and current coefficient, respectively. According to the spectral domain technique [11], the $m$-th ($m = 1, ..., M$) *expansion voltage* can be approximated as a truncated Floquet series

$$V_m(\theta, \varphi, f; \mathbf{t}, \boldsymbol{\varepsilon}, \mathbf{d}) \approx \frac{1}{\mathcal{A}} \sum_{p=-P}^{P} \sum_{q=-Q}^{Q}$$
$$\left\{ G_{zx} \left[ k_x^{p,q}(\theta,\varphi), k_y^{p,q}(\theta,\varphi) \right] \hat{\mathbf{x}} + \right.$$
$$\left. + G_{zy} \left[ k_x^{p,q}(\theta,\varphi), k_y^{p,q}(\theta,\varphi) \right] \hat{\mathbf{y}} \right\} \cdot$$
$$\cdot \mathbf{J}_m^* \left[ k_x^{p,q}(\theta,\varphi), k_y^{p,q}(\theta,\varphi) \right]$$
$$\exp\left[ jk_x^{p,q}(\theta,\varphi) d_5 \right] \exp\left[ jk_y^{p,q}(\theta,\varphi) d_6 \right] \quad (8)$$

[1]Following [11], periodic array conditions are considered hereinafter. Nevertheless, array truncation effects may be taken into account by suitably extending and generalizing (6).

where $\cdot^*$ stands for the complex conjugate, $\mathcal{A} \triangleq |\mathbf{w}_1 \times \mathbf{w}_2|$ is the area of the unit cell of the array lattice, $k_x^{p,q}(\theta,\varphi)$ $[k_x^{p,q}(\theta,\varphi) \triangleq p \times \nu_{1,x} + q \times \nu_{2,x} + \frac{2\pi}{\lambda}\sin(\theta)\cos(\varphi)]$ and $k_y^{p,q}(\theta,\varphi)[k_y^{p,q}(\theta,\varphi) \triangleq p\times\nu_{1,y}+q\times\nu_{2,y}+\frac{2\pi}{\lambda}\sin(\theta)\sin(\varphi)]$ are the spectral domain variables, $\lambda$ is the free-space wavelength at the frequency $f$, $\boldsymbol{\nu}_1$ ($\boldsymbol{\nu}_1 \triangleq \nu_{1,x}\widehat{\mathbf{x}} + \nu_{1,y}\widehat{\mathbf{y}} = 2\pi\frac{\mathbf{w}_2\times\widehat{\mathbf{z}}}{\mathbf{w}_1\cdot\mathbf{w}_2\times\widehat{\mathbf{z}}}$) and $\boldsymbol{\nu}_2$ ($\boldsymbol{\nu}_2 \triangleq \nu_{2,x}\widehat{\mathbf{x}} + \nu_{2,y}\widehat{\mathbf{y}} = 2\pi\frac{\widehat{\mathbf{z}}\times\mathbf{w}_1}{\mathbf{w}_1\cdot\mathbf{w}_2\times\widehat{\mathbf{z}}}$) are the *reciprocal lattice vectors*, $P$ and $Q$ are the truncation indexes of the Floquet modes. Moreover, $\mathbf{J}_m$ ($m = 1,...,M$) in (8) is the spatial Fourier transform of the $m$-th modal basis function $\boldsymbol{\chi}_m(x,y)$ (see the Appendix)

$$\mathbf{J}_m\left[k_x^{p,q}(\theta,\varphi),k_y^{p,q}(\theta,\varphi)\right] = \int_{-\frac{d_3}{2}}^{\frac{d_3}{2}}\int_{-\frac{d_4}{2}}^{\frac{d_4}{2}}\{\boldsymbol{\chi}_m(x,y) \\ \exp\left[jk_x^{p,q}(\theta,\varphi)x\right]\exp\left[jk_y^{p,q}(\theta,\varphi)y\right]\}\,dxdy \quad (9)$$

and the unknown terms $\{C_m(\cdot); m = 1,...,M\}$ in (7) are determined by solving the following system of $M$ linear equations

$$\sum_{m=1}^{M}\sigma_{mh}(\theta,\varphi)\,C_m(\theta,\varphi,f;\mathbf{t},\boldsymbol{\varepsilon},\mathbf{d}) = U_h(\theta,\varphi),\\ h = 1,\dots,M \quad (10)$$

where $U_h(\theta,\varphi)$ is the $h$-th ($h = 1,...,M$) known *test voltage term*

$$U_h(\theta,\varphi) = \frac{1}{\mathcal{A}}\sum_{p=-P}^{P}\sum_{q=-Q}^{Q}\mathbf{J}_h\left[k_x^{p,q}(\theta,\varphi),k_y^{p,q}(\theta,\varphi)\right]\cdot \\ \cdot\left[G_{xz}\left[k_x^{p,q}(\theta,\varphi),k_y^{p,q}(\theta,\varphi)\right]\widehat{\mathbf{x}}+ \right. \\ \left. +G_{yz}\left[k_x^{p,q}(\theta,\varphi),k_y^{p,q}(\theta,\varphi)\right]\widehat{\mathbf{y}}\right] \\ \exp\left[jk_x^{p,q}(\theta,\varphi)d_5\right]\exp\left[jk_y^{p,q}(\theta,\varphi)d_6\right], \quad (11)$$

while the $(m,h)$-th ($m = 1,...,M$; $h = 1,...,M$) impedance matrix term $\sigma_{mh}(\theta,\varphi)$ is given by

$$\sigma_{mh}(\theta,\varphi) = -\frac{1}{\mathcal{A}}\sum_{p=-P}^{P}\sum_{q=-Q}^{Q} \\ \mathbf{J}_m\left[k_x^{p,q}(\theta,\varphi),k_y^{p,q}(\theta,\varphi)\right]\cdot \\ \cdot\overline{\overline{\mathcal{G}}}\left[k_x^{p,q}(\theta,\varphi),k_y^{p,q}(\theta,\varphi)\right]\cdot \\ \cdot\mathbf{J}_h^*\left[k_x^{p,q}(\theta,\varphi),k_y^{p,q}(\theta,\varphi)\right] \quad (12)$$

where

$$\overline{\overline{\mathcal{G}}}\left[k_x^{p,q}(\theta,\varphi),k_y^{p,q}(\theta,\varphi)\right] \triangleq \\ \left[\begin{array}{c} G_{xx}\left[k_x^{p,q}(\theta,\varphi),k_y^{p,q}(\theta,\varphi)\right] \\ G_{yx}\left[k_x^{p,q}(\theta,\varphi),k_y^{p,q}(\theta,\varphi)\right] \\ G_{xy}\left[k_x^{p,q}(\theta,\varphi),k_y^{p,q}(\theta,\varphi)\right] \\ G_{yy}\left[k_x^{p,q}(\theta,\varphi),k_y^{p,q}(\theta,\varphi)\right] \end{array}\right] \quad (13)$$

is the spectral domain Green's dyad for the $x$-$y$ field components.

By substituting (7) in (6), the *ARL* of a *PPAW* is yielded once all the Green's function terms for the components of the electric field [i.e., $G_{zx}[\cdot]$ and $G_{zy}[\cdot]$ in (8), $G_{yz}[\cdot]$ and $G_{yz}[\cdot]$ in (11), and $\overline{\overline{\mathcal{G}}}[\cdot]$ in (13)] have been computed (see the Appendix).

It is worth pointing out that, thanks to the semi-analytical nature of (6), the implementation of the *electromagnetic modeling* block for the *ARL* computation entails a numerical efficiency higher than that from using "bare" full-wave solutions. Indeed, the transforms in (9) can be computed in closed form, thus avoiding expensive numerical integrations. Moreover, the proposed derivation turns out to be a generalization of the approach in [11] since it extends the calculation of the *ARL* to uniaxially anisotropic materials (see the Appendix). As for the generality of the arising *SbD*-based synthesis approach, more complex printed antenna geometries (e.g., featuring stacked metallic elements and/or arbitrary contours) and/or alternative cost function definitions, beyond the scope of the current work, may be handled within the same framework by suitably customizing the *DoF* definition, the semi-analytical modeling approach, and the Green's function terms, but without altering the skeleton and the main steps of the whole scheme.

## IV. NUMERICAL VALIDATION

The objective of this section is twofold. First, it is aimed at illustrating the features of the proposed method in synthesizing low-profile *WAIM* superstrates to enhance the scan capabilities of *PPA*s. The second task is to provide suitable guidelines for a successful and reliable application of the *SbD*-based technique at hand. Towards this end, a set of representative examples dealing with different lattices, substrates, and antenna layouts is proposed and discussed in the following. As for the setup of the elementary building blocks of the *SbD* design strategy, the guidelines in [13][26] and [11] have been adopted and the calibration parameters have been set as follows: $R = 10$, $\zeta_1 = 0.4$, $\zeta_2 = \zeta_3 = 2.0$, $\omega = 30$, $\mathcal{C} = 10^{-4}$, $K = 200$, and $P = Q = 60$.

As a first illustrative proof-of-concept example, let us consider the synthesis of a $L = 2$-layers *WAIM* operating at a single frequency (i.e., $f_{min} = f_{max} = 10$ GHz) and coating an array arrangement [11] composed by rectangular patches of width $d_3 = 1.185 \times 10^{-2}$ [m] and length $d_4 = 9.06 \times 10^{-3}$ [m], arranged according to a half-wavelength square lattice [i.e., $d_7 = d_{10} = 1.5 \times 10^{-2}$ [m], $d_8 = d_9 = 0.0$ [m] - Fig. 2(a)], and printed over a substrate having thickness $d_1 = 1.575 \times 10^{-3}$ [m] and relative permittivity $\varepsilon_{sub} = 2.2$, while the feeding point is located at $(d_5,d_6) = (2.298\times10^{-3},5.925\times10^{-3})$ [m]. Moreover, the feasibility sets have been set to $\mathcal{T} \triangleq \{t_l \in [0.0,0.5\lambda]; l = 1,..,L\}$ and $\mathcal{E} \triangleq \{\varepsilon_l \in [1,30]; l = 1,..,L\}$.

By comparing the plot of the active transmission coefficient (*ATC*) [13]

$$|\tau(\theta,\varphi,f;\mathbf{t},\boldsymbol{\varepsilon},\mathbf{d})|^2 \triangleq 1 - [ARL(\theta,\varphi,f;\mathbf{t},\boldsymbol{\varepsilon},\mathbf{d})]^{-1} \quad (14)$$

of the *PPA* with and without the synthesized low-profile (i.e., $\sum_{l=1}^{L} t_l^{opt} \approx 0.61\lambda$ - Tab. I) *WAIM* composed by an isotropic non-negative material (i.e., $\varepsilon_l^{(\alpha)} \leftarrow \varepsilon_l$, $\alpha \in \{xx,yy,zz\}$), it turns out that the *SbD*-designed *PPAW* significantly improves the scanning performance of the original *PPA* since $|\tau(\theta,\varphi)|_{SbD}^2 \geq 90\%$ when $\theta \in [0,62.5]$ [deg], while the standard layout verifies the condition $|\tau(\theta,\varphi)|_{NoWAIM}^2 \geq 90\%$ only within the angular range $\theta \in [0,40]$ [deg] [Fig. 4(a) vs. Fig. 4(b)]. Although the cost function at hand (2) sets an integral constraint on the whole scanning range, the power transfer capabilities of the *PPA* are generally enhanced in every angular direction. For instance, let us analyze the plots of the *ATC* in the $\varphi = \{0, 45, 90\}$ [deg] planes ($\varphi = 0$ being the E-plane) [Fig. 4(c)]. Unless a narrow blind spot at $\theta \approx 65$ [deg] along the $\varphi = 0$ [deg] and $\varphi = 90$ [deg] cuts, which is potentially related to surface wave resonances arising in the presence of the designed coating[2], the exploitation of a *WAIM* layer upgrades the *PPA* performance along each scanning direction.

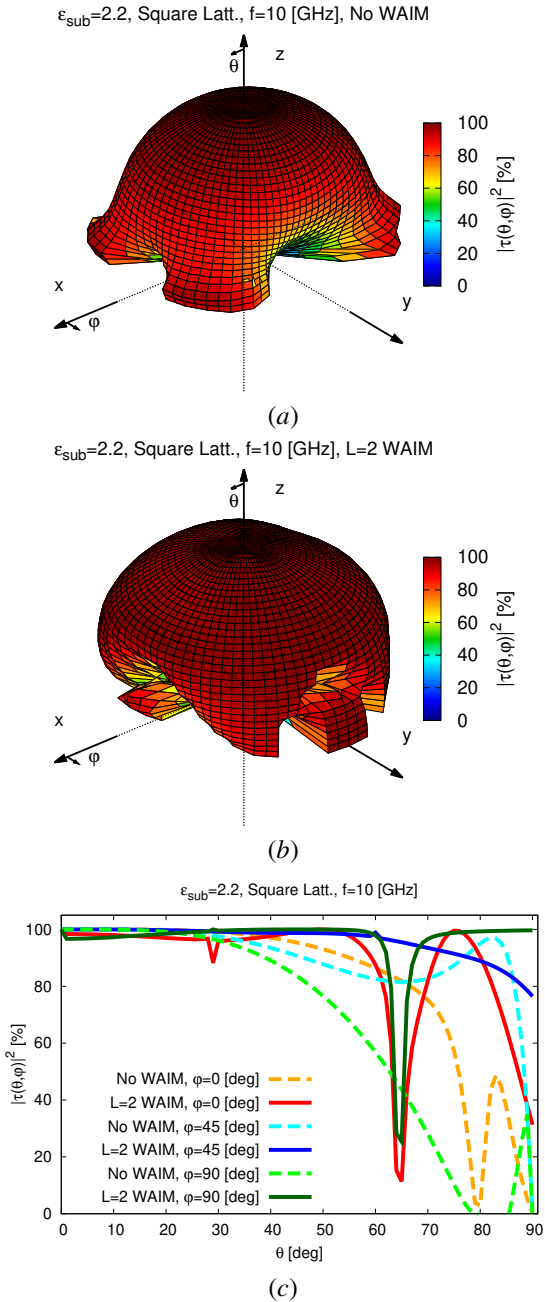

Figure 4. *Single-Frequency Design* ($f = 10$ GHz, Square lattice, $\varepsilon_{sub} = 2.2$, Isotropic *WAIM*, $L = 2$). Plots of $|\tau(\theta, \varphi)|^2$ of (*a*) the uncoated array, (*b*) the *WAIM*-coated array as well as (*c*) the corresponding cuts along the $\varphi = 0$ [deg] and the $\varphi = 90$ [deg] planes.

As for the effectiveness of the *SbD* search process, the plot of the normalized cost function $\Psi_{norm}^{(k)}$

$$\Psi_{norm}^{(k)} \triangleq \frac{\overline{\Psi}^{(k)}}{\Psi^{(No\,WAIM)}} \qquad (15)$$

$\Psi^{(No\,WAIM)}$ being the cost function value for the *PPA* without

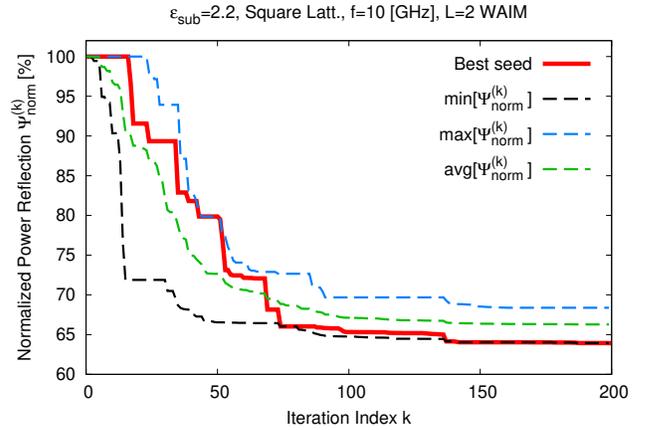

Figure 5. *Single-Frequency Design* ($f = 10$ GHz, Square lattice, $\varepsilon_{sub} = 2.2$, Isotropic *WAIM*, $L = 2$). Behaviour of $\Psi_{norm}^{(k)}$ versus the *SbD* iteration number and associated statistics for different random initializations.

coating (i.e., $\Psi^{(No\,WAIM)} \triangleq \Psi(\{\varnothing\}, \{\varnothing\}, \mathbf{d})$) shows that (*i*) the power reflection is considerably reduced with respect to the original *PPA* in few iterations (e.g., $\Psi_{norm}^{(k)}\big]_{k \approx 70} - \Psi_{norm}^{(0)} \approx -30\%$ - Fig. 5) and that a further mitigation of the power losses is possible thanks to the exploration capabilities of the *PSO*-based global optimization block (e.g., $\Delta\Psi \triangleq \frac{\overline{\Psi}^{(K)} - \Psi^{(No\,WAIM)}}{\Psi^{(No\,WAIM)}} \approx -36.07\%$ - Tab. I); (*ii*) similar convergence performance are achieved regardless of the initialization, as it can be noticed by analyzing the cost function statistics over several random runs (Fig. 5); (*iii*) the occasional "step-like" $\Psi_{norm}^{(k)}$ evolution is caused by the stochastic nature of the implemented *solution space exploration* strategy [26][27], which is not based on a gradient exploration technique (Sect. III). In order to point out also the numerical efficiency of the *SbD* procedure, the values of computational indexes ($\Delta t$ - total time; $\Delta t^{SSE}$- *solution space exploration*; $\Delta t^{PL}$ - *physical linkage*; $\Delta t^{EM}$ - *electromagnetic modeling*) are reported in Tab. I.[3]Despite the quite complex structure featuring a multi-layer superstrate printed on a probe-fed patch antenna array and the large values of the truncation indexes for the Floquet-modes expansion (i.e., $P = Q = 60$, while $P = Q = 4$ are enough for open-ended waveguide arrays [13]), the entire design process requires less than 9 hours to complete (Tab. I). Moreover, it is worth remarking that the overall design time is dominated by $\Delta t^{EM}$ (i.e., $\frac{\Delta t^{EM}}{\Delta t} \approx 99.5\%$ - Tab. I). This result shows the intrinsically challenging nature of the *electromagnetic modeling* problem (Sect. III).

The second numerical experiment is concerned with *PPAs* arranged on non-square lattices. Dealing with a equilateral triangle as the unit cell [i.e., $d_7 = 1.5 \times 10^{-2}$ [m], $d_8 = 0.0$ [m], $d_9 = 1.3 \times 10^{-2}$ [m], $d_8 = 7.5 \times 10^{-3}$ [m] - Fig. 2(*b*)] for the array grid, but keeping the same rectangular elementary radiator of the first example, the values of $\tau(\theta, \varphi)$ from the "*No WAIM*" layout [Fig. 6(*a*)] and from the *SbD PPAW* [Fig.

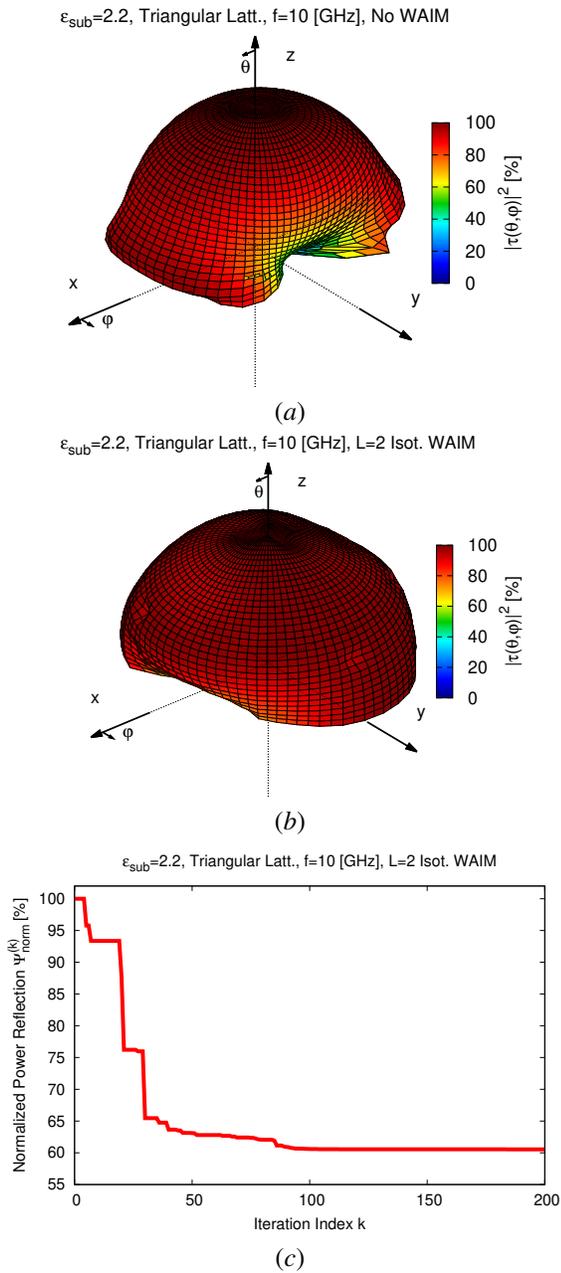

Figure 6.  *Single-Frequency Design* ($f = 10$ GHz, Triangular lattice, $\varepsilon_{sub} = 2.2$, Isotropic *WAIM*, $L = 2$). Plots of $|\tau(\theta, \varphi)|^2$ of (*a*) the uncoated array and (*b*) the *WAIM*-coated array along with (*c*) the evolution of $\Psi_{norm}^{(k)}$ versus the *SbD* iteration number.

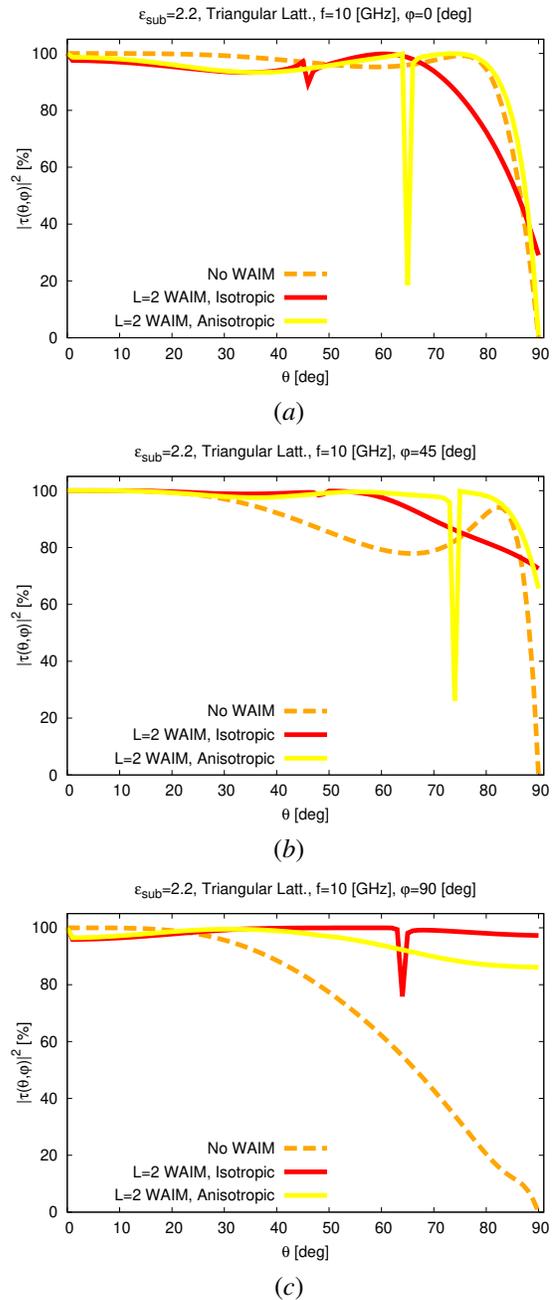

Figure 7.  *Single-Frequency Design* ($f = 10$ GHz, Triangular lattice, $\varepsilon_{sub} = 2.2$, $L = 2$). Plots of $|\tau(\theta, \varphi)|^2$ of the uncoated and the *WAIM*-coated array layouts along (*a*) the $\varphi = 0$ [deg] plane, (*b*) the $\varphi = 45$ [deg] plane, and (*c*) the $\varphi = 90$ [deg] plane.

6(*b*)] indicate that the *ARL* is still improved regardless of the array arrangement [Fig. 6(*b*) vs. Fig. 6(*a*)]. Quantitatively, the power loss is mitigated for an amount of $\Delta\Psi \approx -39.46\%$ (Tab. I) and the values of the *ATC* in the $\varphi = 0$ [deg] plane [Fig. 7(*a*)], the $\varphi = 45$ [deg] plane [Fig. 7(*b*)], and the $\varphi = 90$ [deg] plane [Fig. 7(*c*)] indicate that the *WAIM* layer allows the array to reach a $> 90\%$ power efficiency within the scan range $\theta \in [0, 64]$ [deg] in the worst-case plane [Fig. 7(*c*)], as well. Otherwise, the original *PPA* yields a similar efficiency only when $\theta \le 38$ [deg], where it slightly overcomes the *WAIM*-coated solution [Fig. 7(*c*)]. Such a result is an consequence of the adopted cost function (2), which aims at mitigating the *average* power loss across all angles. To avoid such an effect, an unequal angular weighting of the *ARL* in (2) could be adopted to ensure the performance of the *PPAW* array is at least as good as that in the "No WAIM" case in the not far-scanning angles. Moreover, it is worth noticing that the resulting *PPAW* provides a value of $|\tau(\theta, \varphi)|^2_{SbD} \ge 70\%$ when $\theta = 90$ [deg] in the diagonal steering plane [i.e., $\varphi = 45$ [deg] - Fig. 7(*b*)], thus the possibility to perform end-fire power transfer along this critical direction by exploiting the power delivered to the surface waves along the substrate layers, even though its transformation in radiated field from a truncated array is generally not trivial. This feature is confirmed by the iterative evolution of $\Psi_{norm}^{(k)}$ [Fig. 6(c)]. As it can be observed, a 30% power loss mitigation needs only $k = 30$ *SbD* iterations [i.e., $\left. \Psi_{norm}^{(k)} \right|_{k=30} - \Psi_{norm}^{(0)} \approx -30\%$]. It is worth remarking that the performance of the obtained designs turn out robust also to non-trivial manufacturing tolerances in the layer thickness (Fig. 8). To assess such a property, the *SbD PPAW* design in Fig. 6(b) has been numerically simulated assuming a $\pm 10\%$ fabrication tolerance (Fig. 8). The plots of $|\tau(\theta, \varphi)|^2$ for the "exact" and "perturbed" configurations demonstrate that, despite the considerable fabrication tolerances, the qualitative and quantitative effectiveness of the designed *WAIM* is confirmed in all the angular cuts (Fig. 8). Concerning the computational metrics, the values in Tab. I indicate that the efficiency of the synthesis process does not significantly depend on the lattice shape (i.e., $\frac{\Delta t^{triangular}}{\Delta t^{square}} \approx 0.99$ - Tab. I), thanks to the semi-analytic nature of the *electromagnetic modeling* block (see Sect. III), and that the overall design time is once again dominated by the inherent complexity of such a phase (Tab. I).

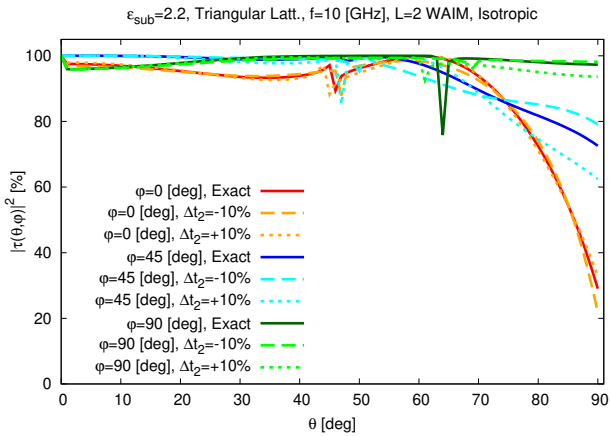

Figure 8. *Single-Frequency Design* ($f = 10$ GHz, Triangular lattice, $\varepsilon_{sub} = 2.2$, $L = 2$, Isotropic WAIM). Plots of $|\tau(\theta, \varphi)|^2$ along the $\varphi \in \{0, 45, 90\}$ [deg] planes with exact thickness or $\pm 10\%$ manufacturing tolerances.

The capability of the *SbD*-based design technique to handle more complex constraints and materials as well as to deduce different trade-offs, in terms of *WAIM* complexity versus performance, is assessed next. More specifically, uniaxially anisotropic materials have been considered for the same triangular layout of the previous example. As expected, the *ATC* of the anisotropic *PPAW* turns out to be better than that of its corresponding isotropic counterpart as shown by the plots in Fig. 9(a) vs. Fig. 6(b) and quantitatively confirmed by the power loss indexes in Tab. II. Moreover, while the material composing the device is more complex, its structure turns out to be simplified since the thickness of the synthesized *WAIM* layer is further reduced with respect to the (already) low-profile isotropic case ($\frac{\sum_{l=1}^{L} t_l^{opt}|^{aniso}}{\sum_{l=1}^{L} t_l^{opt}|^{iso}} \approx 0.96$ - Tab. I vs. Tab. II). This is not surprising both considering the performance of state-of-the-art anisotropic *WAIM* designs when combined with different antenna elements [13] and also taking into account the pillars of the optimization theory since an equal or better design can be usually found whether a wider (physically admissible) search range is adopted [26][27]. On the other hand, it should be noted that there are similar convergence behaviours [Fig. 9(b) vs. Fig. 6(c)] and final values of $\Delta \Psi$ when using anisotropic or isotropic materials with only a moderate power loss improvement thanks to the more complex materials ($\Delta \Psi^{aniso} \approx -39.75\%$ vs. $\Delta \Psi^{iso} \approx -39.46\%$ - Tab. II vs. Tab. I). Such outcomes seem to indicate that there is no so big advantage in "enlarging" the set of the material properties at the expense of higher costs and a more difficult manufacturing process. Therefore, isotropic *WAIMs* will be analyzed hereinafter taking into account the aim of this research track of providing the readers of suitable insights on the effectiveness of the *SbD* synthesis tool when simple materials/structures are at hand.

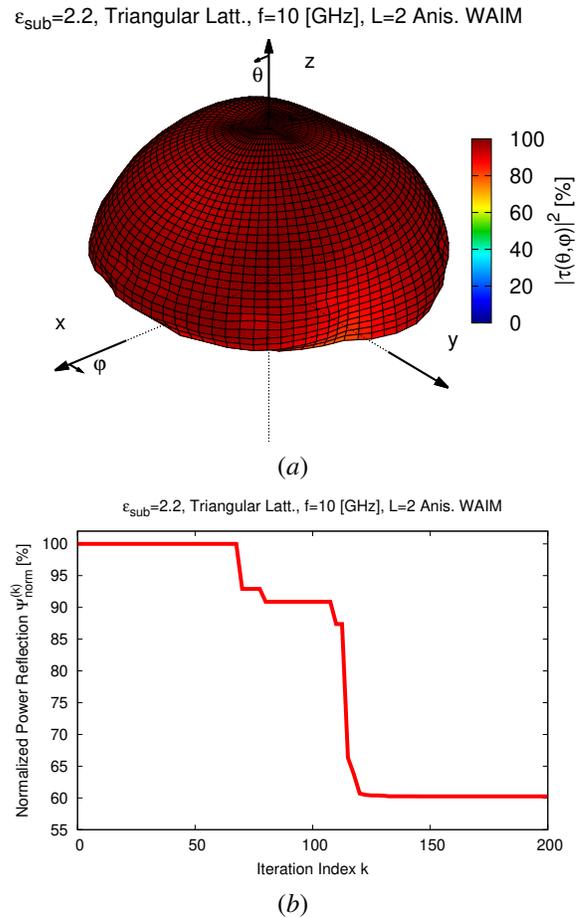

Figure 9. *Single-Frequency Design* ($f = 10$ GHz, Triangular lattice, $\varepsilon_{sub} = 2.2$, Anisotropic *WAIM*, $L = 2$). Plots of (a) $|\tau(\theta, \varphi)|^2$ of the *WAIM*-coated array and (b) evolution of $\Psi_{norm}^{(k)}$ versus the *SbD* iteration number.

Along this line of reasoning, the *SbD*-based *WAIM* design has been evaluated when varying the number of (isotropic) layers of the *PPA* superstrate to keep positive the balance between performance increasing and structure/material simplicity/costs. From the plots of the power loss improvement factor versus $L$ for a square or a triangular array lattice [Fig. 10(a)] and the corresponding descriptors (Tab. III), the following considerations can be drawn: (*i*) single-layer *WAIM*s negligibly outperform "*No WAIM*" *PPA*s [i.e., $\Delta\Psi|_{L=1}^{square} \approx -0.03\%$, $\Delta\Psi|_{L=1}^{triang} \approx -0.04\%$ - Fig. 10(*a*)], the optimized superstrate being close to a free-space fictitious layer ($\varepsilon_1^{opt}|_{L=1} \approx 1.01$, $\varepsilon_1^{opt}|_{L=1}^{triang} \approx 1.03$ - Tab. III); (*ii*) $L = 2$ and $L = 3$ layers solutions considerably reduce the power losses [Fig. 10(*a*)], but there are only slightly benefits in adding more layers besides the second one (i.e., $\Delta\Psi|_{L=3}^{square} - \Delta\Psi|_{L=2}^{square} \approx -0.08\%$, $\Delta\Psi|_{L=3}^{triang} - \Delta\Psi|_{L=2}^{triang} \approx -0.25\%$ - Tab. III vs. Tab. I); (*iii*) the $L = 3$ optimal *WAIM* is composed by two ($l = 1$ and $l = 3$) low-permittivity materials (i.e., $\varepsilon_1|_{L=3}^{square} \approx 1.04$, $\varepsilon_1|_{L=3}^{triang} \approx 1.01$; $\varepsilon_3|_{L=3}^{square} \approx 1.16$, $\varepsilon_3|_{L=3}^{triang} \approx 1.08$ - Tab. III), while the internal one ($l = 2$) features $\varepsilon_2 > 1.5$ media (i.e., $\varepsilon_2|_{L=3}^{square} \approx 1.55$, $\varepsilon_2|_{L=3}^{triang} \approx 2.18$ - Tab. III). It is actually close to a $L = 2$ arrangement added of a fictitious free-space top layer (Tab. III), since $\Delta\Psi|_{L=3} \approx \Delta\Psi|_{L=2}$ regardless of the lattice under analysis [Fig. 10(*a*)]. Interestingly, effective *WAIM*s for wide scanning consist of a low-permittivity first ($l = 1$) layer combined with a superimposed ($l = 2$) higher-contrast one (i.e., $\varepsilon_1|_{L=2}^{square} \approx 1.04$, $\varepsilon_1|_{L=2}^{triang} \approx 1.03$; $\varepsilon_2|_{L=2}^{square} \approx 3.13$, $\varepsilon_2|_{L=2}^{triang} \approx 3.42$ - Tab. I). All previous outcomes are further contextualized in Fig. 10 where the plots of the *ATC* in the $\varphi = 0$ [deg] plane [square lattice - Fig. 10(*b*); triangular lattice - Fig. 10(*c*)] and in the $\varphi = 90$ [deg] plane [square lattice - Fig. 10(*d*); triangular lattice - Fig. 10(*e*)] are reported. The last comment on this example is related to the dependence of the computational load of the *SbD* synthesis on the number of *WAIM* layers. As expected and analogously to the case of anisotropic materials, the more complex *WAIM* structure does not significantly impact on the *CPU* time (i.e., $\frac{\Delta t|_{L=3}^{square}}{\Delta t|_{L=2}^{square}} \approx 1.18$, $\frac{\Delta t|_{L=3}^{triang}}{\Delta t|_{L=2}^{triang}} \approx 1.17$ - Tab. III) once again thanks to the semi-analytical nature of the *electromagnetic modeling block* that does not imply a full-wave modeling (i.e., $\lambda$-dependent discretization) and simulation of the *WAIM* superstrate (Sect. III).

Table I
*Single-Frequency Design* ($f = 10$ GHz, ISOTROPIC *WAIM*, $L = 2$). SOLUTION DESCRIPTORS AND FIGURES OF MERIT.

| *Fig.* | $\varepsilon_{sub}$ | *Latt.* | $t_1^{opt}$ [m] | $t_3^{opt}$ [m] | $\varepsilon_1^{opt}$ | $\varepsilon_3^{opt}$ | $\Delta\Psi$ | $\Delta t$ [s] | $\Delta t^{SSE}$ [s] | $\Delta t^{PL}$ [s] | $\Delta t^{EM}$ [s] |
|---|---|---|---|---|---|---|---|---|---|---|---|
| 4 | 2.2 | Sq. | $4.65 \times 10^{-3}$ | $1.34 \times 10^{-2}$ | 1.04 | 3.13 | $-36.07\%$ | $3.08 \times 10^4$ | $1.50 \times 10^2$ | $1.52 \times 10^1$ | $3.06 \times 10^4$ |
| 6 | 2.2 | Tri. | $3.54 \times 10^{-3}$ | $1.33 \times 10^{-2}$ | 1.03 | 3.42 | $-39.46\%$ | $3.05 \times 10^4$ | $1.35 \times 10^2$ | $1.49 \times 10^1$ | $3.03 \times 10^4$ |
| 11($c$) | 12.8 | Sq. | $9.95 \times 10^{-4}$ | $9.87 \times 10^{-4}$ | 29.28 | 23.84 | $-92.67\%$ | $3.06 \times 10^4$ | $1.42 \times 10^2$ | $1.51 \times 10^1$ | $3.04 \times 10^4$ |
| 11($d$) | 12.8 | Tri. | $9.93 \times 10^{-4}$ | $9.84 \times 10^{-4}$ | 29.37 | 23.71 | $-92.74\%$ | $3.31 \times 10^4$ | $1.65 \times 10^2$ | $1.54 \times 10^1$ | $3.29 \times 10^4$ |

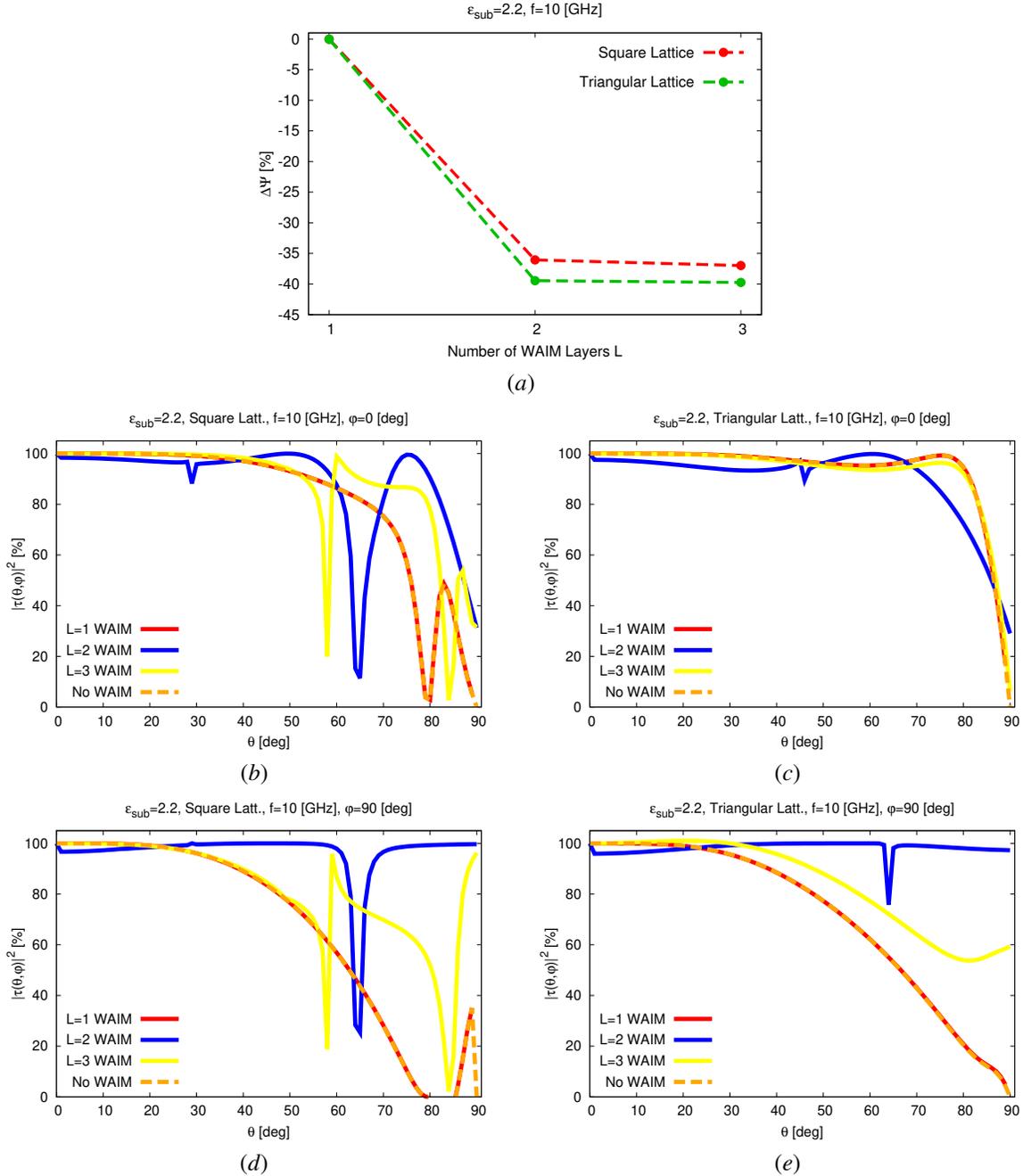

Figure 10.   *Single-Frequency Design* ($f = 10$ GHz, $\varepsilon_{sub} = 2.2$, Isotropic *WAIM*, $L \in [1, 3]$). Behaviour of $\Delta\Psi$ versus $L$ ($a$) and plots of $|\tau(\theta, \varphi)|^2$ of the uncoated and the *WAIM*-coated array layouts along ($b$)($c$) $\varphi = 0$ [deg] and ($d$)($e$) $\varphi = 90$ [deg] planes for ($b$)($d$) a square and ($c$)($e$) a triangular array lattice.

Table II
*Single-Frequency Design* ($f = 10$ GHz, TRIANGULAR LATTICE,
$\varepsilon_{sub} = 2.2$, ANISOTROPIC *WAIM*, $L = 2$). SOLUTION DESCRIPTORS AND
FIGURES OF MERIT.

| Quantity | Value |
|---|---|
| $t_1^{opt}$ [m] | $5.7 \times 10^{-3}$ |
| $t_2^{opt}$ [m] | $1.2 \times 10^{-2}$ |
| $\varepsilon_1^{(xx),opt}$ | 1.02 |
| $\varepsilon_1^{(yy),opt}$ | 1.07 |
| $\varepsilon_1^{(zz),opt}$ | 17.38 |
| $\varepsilon_2^{(xx),opt}$ | 2.89 |
| $\varepsilon_2^{(yy),opt}$ | 29.44 |
| $\varepsilon_2^{(zz),opt}$ | 1.74 |
| $\Delta\Psi$ | $-39.75\%$ |
| $\Delta t$ [s] | $2.71 \times 10^4$ |

While previous examples have dealt with *PPA*s printed on a low permittivity medium (i.e., $\varepsilon_{sub} = 2.2$), the fourth test is concerned with higher permittivity substrates usually adopted for array miniaturization purposes (Fig. 11). Towards this end, the *PPA* setup taken from [11] has been analyzed. It consists of a substrate with thickness $d_1 = 1.80 \times 10^{-3}$ [m] and relative permittivity $\varepsilon_{sub} = 12.8$, while the elementary radiator is a printed patch of width $d_3 = 4.50 \times 10^{-3}$ [m] and length $d_4 = 2.94 \times 10^{-3}$ [m], feed at $(d_5, d_6) = (1.47 \times 10^{-3}, 0.0)$ [m]. Moreover, both the half-wavelength square [i.e., $d_7 = d_{10} = 1.5 \times 10^{-2}$ [m], $d_8 = d_9 = 0.0$ [m] - Fig. 2($a$)] and the equilateral triangular [i.e., $d_7 = 1.5 \times 10^{-2}$ [m], $d_8 = 0.0$ [m], $d_9 = 1.3 \times 10^{-2}$ [m], $d_8 = 7.5 \times 10^{-3}$ [m] - Fig. 2($b$)] lattices have been taken into account. The results of the *SbD* synthesis at $f = 10$ [GHz] (i.e., $f_{\min} = f_{\max} = f$) when setting $L = 2$ are summarized in Fig. 11 (square lattice - left column; triangular lattice - right column). By comparing the arising impedance matching patterns with those of the array without *WAIM* [Figs. 11($a$)-11($b$)], it turns out that the power losses are reduced ( $\Delta\Psi|^{square} \approx -92.67\%$, $\Delta\Psi|^{triang} \approx -92.74\%$ - Tab. I) and the *FoV* is widened. As for this latter, it is worth remarking that in many radar/communication systems the useful scan range (i.e., the *FoV*) is often considered to be limited to the scan blindness angle, and that impedance matching improvements beyond such angle are not usually exploited. Within this respect, the result for the square lattice is that $|\tau(\theta,\varphi)|^2_{SbD} \geq 90\%$ when $\theta \leq 85$ [deg], while $|\tau(\theta,\varphi)|^2_{NoWAIM} \geq 90\%$ only if $\theta \leq 27$ [deg] - Fig. 11($a$), showing a $\approx 58$ [deg] improvement in terms of maximum scan range enabled by the designed *WAIM*. Similarly for the triangular arrangement, $|\tau(\theta,\varphi)|^2_{SbD} \geq 90\%$ when $\theta \leq 90$ [deg] - Fig. 11($b$) versus 35 [deg] scan range only of the standard arrangement [Fig. 11($b$)], corresponding to a $\approx 55$ [deg] *FoV* widening. Once again, regardless of the patch substrate and the geometrical displacement of the array elements, the proposed *SbD* implementation proved to be a computationally efficient ($\Delta t \leq 3.31 \times 10^4$ - Tab. I) and reliable *WAIM* design tool.

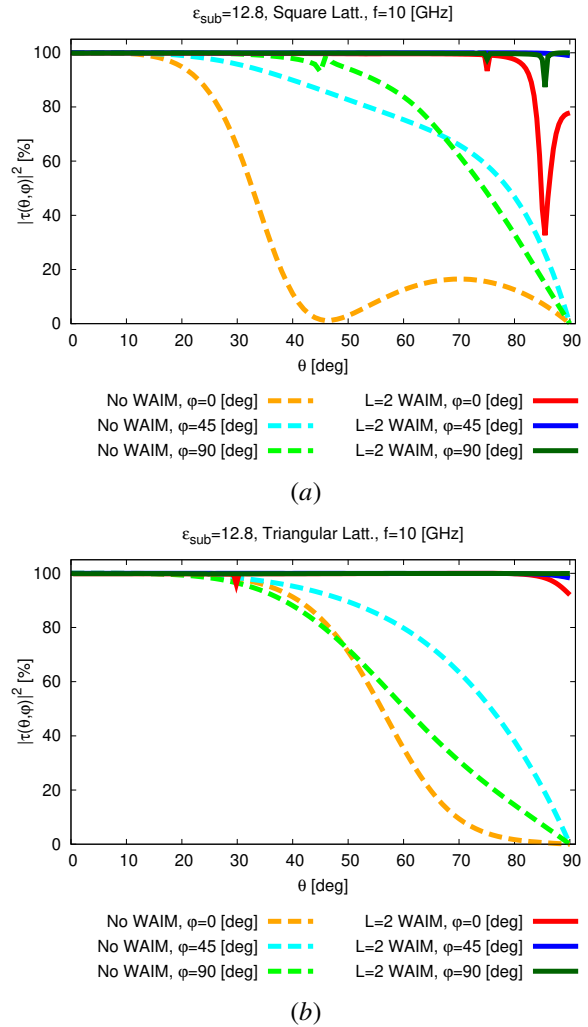

Figure 11. *Single-Frequency Design* ($f = 10$ GHz, $\varepsilon_{sub} = 12.8$, Isotropic *WAIM*, $L = 2$). Plots of $|\tau(\theta, \varphi)|^2$ of the uncoated array and the *WAIM*-coated array along the $\varphi = 0$ [deg], the $\varphi = 45$ [deg], and the $\varphi = 90$ [deg] planes for (*a*) a square and (*b*) a triangular array lattice.

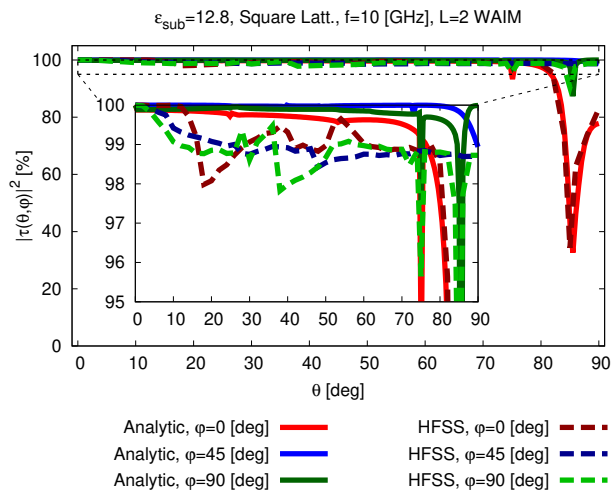

Figure 12. *Single-Frequency Design* ($f = 10$ GHz, $\varepsilon_{sub} = 12.8$, Isotropic *WAIM*, $L = 2$, Square Lattice). Plots of analytically-modeled and *HFSS* simulated $|\tau(\theta, \varphi)|^2$ of the *WAIM*-coated array in the cuts along the $\varphi \in \{0, 45, 90\}$ [deg] planes.

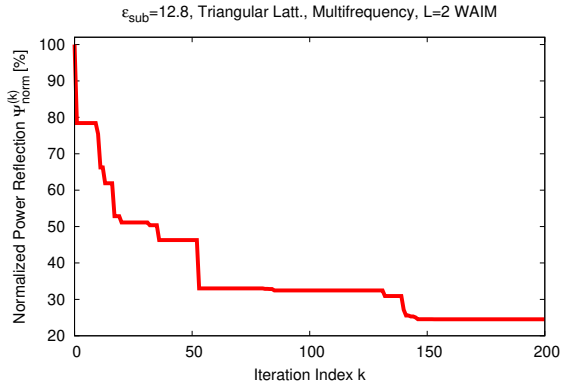

Figure 13.   *Multi-Frequency Design* ($f \in [9, 11]$ GHz, Triangular lattice, $\varepsilon_{sub} = 12.8$, Isotropic *WAIM*, $L = 2$). Behaviour of $\Psi_{norm}^{(k)}$ versus the *SbD* iteration number.

For the sake of completeness, the consistency of the *semi-analytical methodology* developed for the *electromagnetic modeling* block (Sect. III) with full-wave commercial softwares has been assessed next. Towards this end, the optimized $L = 2$ design in Fig. 11($a$) has been simulated using the finite-element solver of Ansys HFSS [34]. The results reported in Fig. 12 show that despite its semi-analytical nature, the proposed analysis strategy affords a satisfactory accuracy in the entire scan range, with minimal *ARL* misalignments (i.e., $\left| |\tau(\theta,\varphi)|^2_{ANA} - |\tau(\theta,\varphi)|^2_{HFSS} \right| < 2\%$ on the average - Fig. 12) with respect to the full-wave simulator regardless of the considered cuts.

The last experiment is devoted to the assessment of the *SbD*-based strategy when wideband/multi-frequency operations are of interest. Towards this end, a *WAIM* superstrate coating the triangularly-arranged layout with higher permittivity substrate (i.e., $\varepsilon_{sub} = 12.8$) has been synthesized for wide scanning within the frequency range $[f_{\min}, f_{\max}] = [9, 11]$ [GHz] (Fig. 13). The evolution of the value of the normalized cost function during the iterative optimization indicates that, also in this case, the power reflection is significantly reduced after few *SbD* iterations (i.e., $\Psi_{norm}^{(k)}\big|_{k=11} - \Psi_{norm}^{(0)} \approx -30\%$ - Fig. 13), the power loss mitigation with respect to the array without *WAIM* being greater than 75% ($\Delta\Psi \approx -75.46\%$ - Tab. IV, Fig. 13) notwithstanding the use of isotropic materials and very low-profile superstrates (i.e., $\sum_{l=1}^{L} t_l^{opt} \approx 0.1\lambda$).

## V. CONCLUSIONS AND FINAL REMARKS

An innovative strategy for the synthesis of low-profile *WAIM* superstrates to enlarge the scanning range of microstrip printed phased arrays has been proposed. By exploiting an instance of the *SbD* paradigm, a task-oriented formulation of the *WAIM* synthesis problem has been presented where a customized global optimization strategy has been combined with an accurate and fast modeling of the array active return loss, which is based on a semi-analytical approach, to enable an effective design of multi-layer structures with both isotropic and anisotropic materials.

The numerical validation has shown that

- the proposed method can be effectively used to design low-profile *WAIM*s for considerably widening the steering region of otherwise narrow-scan antenna arrays regardless of the array substrate [e.g., Fig. 11] and without necessarily recurring to anisotropic or negative permittivity materials (e.g., Tab. I);

- the optimization strategy along with the exploitation of a semi-analytical electromagnetic modeling features a fast converging sampling of the space of physically-admissible solutions and a corresponding non-negligible improvement of the performance just after few iterations [Fig. 6($c$)];

- the method can be seamlessly adopted in the case of single/multiple layers, different antenna layouts for both single and multi-frequency operations (e.g., Fig. 10, Fig. 13).

As for the main methodological advancements with respect to the state-of-the-art, they include (*i*) the derivation of an innovative *SbD*-based *WAIM* design strategy that, unlike [13], is suitable for periodic arrangements of printed antennas and it is based on the exploitation of a Galerkin method in the spectral domain to perform the modeling the array *ARL* in a semi-analytical fashion, while a mode-matching strategy was used in [13], (*ii*) the presentation of a semi-analytical derivation of the array *ARL* when printed layouts are coated with multi-layer dielectric superstrates, then used in the design process instead of previously adopted full-wave modeling the whole device through commercial softwares [18], (*iii*) the analysis and the derivation of suitable guidelines for an effective and reliable use of such a synthesis tool, and (*iv*) the customization and integration of a multi-agent global optimization algorithm within the design process, unlike previous techniques for the design of *WAIMs* for printed antennas based on local search approaches [18].

Table III
*Single-Frequency Design ($f = 10$ GHz, $\varepsilon_{sub} = 2.2$, Isotropic WAIM, $L \in [1,3]$). Solution descriptors and figures of merit.*

| Fig. | L | Latt. | $t_1^{opt}$ [m] | $t_2^{opt}$ [m] | $t_3^{opt}$ [m] | $\varepsilon_1^{opt}$ | $\varepsilon_2^{opt}$ | $\varepsilon_3^{opt}$ | $\Delta\Psi$ | $\Delta t$ [s] |
|---|---|---|---|---|---|---|---|---|---|---|
| 10(b) | 1 | Sq. | $1.00 \times 10^{-3}$ | n.a. | n.a. | 1.01 | n.a. | n.a. | $-0.03\%$ | $2.13 \times 10^4$ |
| 10(c) | 1 | Tri. | $9.97 \times 10^{-4}$ | n.a. | n.a. | 1.03 | n.a. | n.a. | $-0.04\%$ | $2.25 \times 10^4$ |
| 10(d) | 3 | Sq. | $6.61 \times 10^{-3}$ | $1.50 \times 10^{-2}$ | $7.50 \times 10^{-3}$ | 1.04 | 1.55 | 1.16 | $-36.15\%$ | $2.52 \times 10^{-4}$ |
| 10(e) | 3 | Tri. | $9.94 \times 10^{-4}$ | $1.03 \times 10^{-3}$ | $9.91 \times 10^{-4}$ | 1.01 | 2.18 | 1.08 | $-39.71\%$ | $2.64 \times 10^{-4}$ |

Future works, beyond the scope of this paper, will include the generalization of this approach to multi-layered antennas, possibly including parasitic radiators, as well as to layouts including printed antennas with non-canonical contours to support ultra-wideband or multi-band operations. Moreover, the possibility to account for further performance indicators in the design process (such as gain fluctuations and efficiency, in order to mitigate the possible confinement of energy between the dielectric layers even if the *ARL* is low) as well as to account for array truncation effects for small aperture scenarios will be the subject of future investigations.

## Acknowledgements

The authors would like to thank Dr. Luca Manica for the so kind and very useful discussions and comments. A. Massa wishes to thank E. Vico for her never-ending inspiration, support, guidance, and help.

## Appendix

*Expression of $\chi_m(x, y)$, $m = 1, ..., M$ in (9)*

According to the numerical analysis on the convergence issues illustrated in [11], a reliable result for the numerical evaluation of the *ARL* can be obtained when the following $M = 6$ modes are employed

$$\chi_m(x, y) = \begin{cases} \sin\left[\frac{\pi(2m-1)}{d_4}\left(x + \frac{d_4}{2}\right)\right]\hat{x} & m \in [1,4] \\ \sin\left[\frac{\pi(m-4)}{d_3}\left(y + \frac{d_3}{2}\right)\right]\hat{y} & m = 2m \in [5,6] \end{cases} \quad (16)$$

$m = 1$ being the dominant mode of the cavity model.

*Computation of the Green's Function Terms in (8) and (13)*

By generalizing the guidelines in [11] to the case of uniaxially anisotropic materials, the Green's function terms for the components of the electric field [i.e., $G_{zx}[\cdot]$ and $G_{zy}[\cdot]$ in (8), $G_{yz}[\cdot]$ and $G_{yz}[\cdot]$ in (11), and $\overline{\overline{G}}[\cdot]$ in (13)] are

$$G_{xx}\left[k_x^{p,q}(\theta,\varphi), k_y^{p,q}(\theta,\varphi)\right] = \frac{jk_x^{p,q}(\theta,\varphi)\beta_{sub}a_{sub}(1-\Gamma_{sub})}{2\pi f\varepsilon_0\varepsilon_{sub}} + -jk_y^{p,q}(\theta,\varphi)b_{sub}(1+\Phi_{sub}) \quad (17)$$

$$G_{yx}\left[k_x^{p,q}(\theta,\varphi), k_y^{p,q}(\theta,\varphi)\right] = \frac{jk_y^{p,q}(\theta,\varphi)\beta_{sub}a_{sub}(1-\Gamma_{sub})}{2\pi f\varepsilon_0\varepsilon_{sub}} + +jk_x^{p,q}(\theta,\varphi)b_{sub}(1+\Phi_{sub}) \quad (18)$$

$$G_{yy}\left[k_x^{p,q}(\theta,\varphi), k_y^{p,q}(\theta,\varphi)\right] = \frac{jk_y^{p,q}(\theta,\varphi)\beta_{sub}a_{sub}(1-\Gamma_{sub})}{2\pi f\varepsilon_0\varepsilon_{sub}} + -jk_x^{p,q}(\theta,\varphi)b_{sub}(1+\Phi_{sub}) \quad (19)$$

and

$$G_{zx}\left[k_x^{p,q}(\theta,\varphi), k_y^{p,q}(\theta,\varphi)\right] = \frac{\gamma a_{sub}}{j\omega\varepsilon_0\varepsilon_{z,1,1}}\left[\frac{\exp(-j\beta_{sub}d_1)-1}{j\beta_{sub}} + -\frac{\exp(j\beta_{sub}d_1)-1}{j\beta_{sub}}\Gamma_{sub}\right]. \quad (20)$$

where the expression for $\gamma$ is $\gamma \triangleq \left\{\left[k_x^{p,q}(\theta,\varphi)\right]^2 + \left[k_y^{p,q}(\theta,\varphi)\right]^2\right\}$, the substrate propagation coefficient is

$$\beta_{sub} = \sqrt{\frac{2\pi\varepsilon_{sub}}{\lambda} - \gamma}, \quad (21)$$

the reflection coefficients in the antenna substrate are $\Gamma_{sub} = \exp(-2j\beta_{sub}d_1)$, $\Phi_{sub} = -\exp(-2j\beta_{sub}d_1)$, the boundary unknowns are equal to

$$a_{sub} \triangleq \frac{-k_x^{p,q}(\theta,\varphi)(1-\Gamma_1)\frac{\beta_1^{(x)}}{\varepsilon_1^{(x)}}}{A \times \gamma}, \quad (22)$$

$$b_{sub} \triangleq \frac{\omega\mu_0 k_y^{p,q}(\theta,\varphi)(1+\Phi_1)}{B \times \gamma}, \quad (23)$$

the propagation coefficient in the $l$-th ($l = 1, ..., L$) layer is $\beta_l^{(\alpha)} = \sqrt{\frac{2\pi\varepsilon_l^{(\alpha)}}{\lambda} - \gamma}$, $\alpha \in \{xx, yy, zz\}$, and

$$A = j\left[\frac{\beta_{sub}}{\varepsilon_{sub}}(1+\Gamma_1)(1-\Gamma_{sub}) + \frac{\beta_1^{(x)}}{\varepsilon_1^{(x)}}(1+\Gamma_{sub})(1-\Gamma_1)\right] \quad (24)$$

$$B = j\left[\beta_{sub}(1+\Phi_1)(1-\Phi_{sub}) + \beta_1^{(x)}(1+\Phi_{sub})(1-\Phi_1)\right]. \quad (25)$$

Moreover, the reflection coefficients $\Gamma_l$ and $\Phi_l$ ($l = 1, .., L$) can be yielded from the following iterative procedure

$$\Gamma_l = \begin{cases} \exp\left(-2j\beta_l^{(x)}t_l\right)\frac{Z_l - Z_{l+1}^T}{Z_l + Z_{l+1}^T} & l = 1, ..., L-1 \\ 0 & l = L \end{cases} \quad (26)$$

$$\Phi_l = \begin{cases} \exp\left(-2j\beta_l^{(x)}t_l\right)\frac{Y_l - Y_{l+1}^T}{Y_l + Y_{l+1}^T} & l = 1, ..., L-1 \\ 0 & l = L \end{cases}, \quad (27)$$

$Z_l^T$ and $Y_l^T$ being and the impedance ($Z_l^T \triangleq \frac{1-\Gamma_l}{1+\Gamma_l}Z_l$) and the admittance ($Y_l^T \triangleq \frac{1-\Phi_l}{1+\Phi_l}Y_l$) for the electric and the magnetic vector potentials given by [12]

$$Z_l = \frac{\beta_l^{(x)}}{\omega\varepsilon_0\varepsilon_l^{(x)}} \quad (28)$$

$$Y_l = \frac{\beta_l^{(x)}}{\omega\mu_0}. \quad (29)$$

Finally, the other Green's function terms can be inferred thanks to their intrinsic symmetries [11]: $G_{xy}\left[k_x^{p,q}\left(\theta,\varphi\right), k_y^{p,q}\left(\theta,\varphi\right)\right] = G_{yx}\left[k_y^{p,q}\left(\theta,\varphi\right), k_x^{p,q}\left(\theta,\varphi\right)\right]$, $G_{xz}\left[k_x^{p,q}\left(\theta,\varphi\right), k_y^{p,q}\left(\theta,\varphi\right)\right] = -G_{zx}\left[k_x^{p,q}\left(\theta,\varphi\right), k_y^{p,q}\left(\theta,\varphi\right)\right]$, $G_{zy}\left[k_x^{p,q}\left(\theta,\varphi\right), k_y^{p,q}\left(\theta,\varphi\right)\right] = G_{zz}\left[k_y^{p,q}\left(\theta,\varphi\right), k_x^{p,q}\left(\theta,\varphi\right)\right]$, and $G_{yz}\left[k_x^{p,q}\left(\theta,\varphi\right), k_y^{p,q}\left(\theta,\varphi\right)\right] = -G_{zy}\left[k_x^{p,q}\left(\theta,\varphi\right), k_y^{p,q}\left(\theta,\varphi\right)\right]$.